\documentclass[reprint, groupedaddress, amsmath, amssymb, aps]{revtex4-1}

\usepackage{newtxtext}
\usepackage{newtxmath}

\usepackage{graphicx}
\usepackage{dcolumn}
\usepackage{bm}
\usepackage{hyperref}
\hypersetup{
    colorlinks=true,
    linkcolor=blue,
    filecolor=blue,
    urlcolor=blue,
    citecolor = blue
}
\usepackage{xcolor}

\newcommand{\offeq}{\stackrel{\text{\tiny off}}{=}}
\newcommand{\offprop}{\stackrel{\text{\tiny off}}{\propto}}

\begin{document}

\preprint{APS/123-QED}

\title{Analytical theory of pyrochlore cooperative paramagnets}

\author{Akshat Pandey$^{1}$}
\author{Roderich Moessner$^{2}$}
\author{Claudio Castelnovo$^{1}$}
\affiliation{$^1$TCM Group, Cavendish Laboratory, University of Cambridge, Cambridge CB3 0HE, United Kingdom}
\affiliation{$^2$Max Planck Institute for the Physics of Complex Systems, N\"{o}thnitzer Strasse 38, 01187 Dresden, Germany}

\begin{abstract}
The pyrochlore lattice is associated with several potential and actual spin liquid phases as a result of its strong geometric frustration. At finite temperature, these can exhibit an unusually broad cross-over regime to a conventional paramagnet. Here, we study this regime analytically by showing how a single-tetrahedron Hamiltonian can extrapolate beyond the first term of a high-temperature expansion and yield insights into the build-up of correlations. We discuss how this unusual behaviour is brought about by the structure of the eigenspaces of the coupling matrix. 
Further interesting behaviour can appear for parameter values located near phase transitions: we find coexistence of $(111)$ rods and $(220)$ peaks in the structure factor, as observed in neutron scattering experiments on Yb$_2$Ti$_2$O$_7$. 
\end{abstract}

\maketitle
%
%

\section{Introduction} 

Rare-earth pyrochlores have been central to the study of frustrated magnetism, hosting exotic classical and quantum many-body physics, notably emergent gauge fields and putative quantum spin liquid phases~\cite{Castelnovo2012, Gingras2014, Rau2019}. A range of these systems is well-described by nearest-neighbour exchange Hamiltonians which, using the symmetries of the pyrochlore lattice, are completely specified by four real coupling constants~\cite{Curnoe2007, McClarty2009}.

While a complete quantum many-body investigation of these systems remains beyond reach, much can be learned from a classical picture where spin operators are replaced by classical fixed-length vectors. Yan \textit{et al.} produced a rather complete description of the classical $T=0$ phases~\cite{Yan2017}: ground states of one tetrahedron are classified according to the irreducible representations of the tetrahedral point group $T_d$ with which their respective order parameters transform. Ground states of the lattice can be built by connecting single-tetrahedron ground states through the ``Lego-brick rules''~\cite{Yan2017}. These ground states capture the low-temperature behaviour of some rare-earth pyrochlore materials.

At finite temperature the behaviour of these systems is somewhat less understood. The elastic structure factor, probed by neutron scattering experiments~\cite{jensen}, often shows a resemblance between large temperatures and a naive average over single-tetrahedron ground states. An example where the resemblance is an exact equality is ``rods'' in the $(111)$ direction in spin ice, as explained in Ref.~\cite{Castelnovo2019}. Such rods do not signify entirely decorrelated $(111)$ planes, but are artefacts of a term in the structure factor which projects out correlations that are parallel to the wavevector (see Eq.~\eqref{ns}), in this case $(111)$. The coexistence of three-dimensional longitudinal correlations and rod features had already been noted by Thompson \textit{et al.} in the context of the pyrochlore compound Yb$_2$Ti$_2$O$_7$~\cite{Thompson2011}. We will provide further examples of this resemblance, as well as a general framework to understand it. 

It is also observed that for some choices of nearest-neighbour couplings $\mathcal{J}$ on the pyrochlore lattice, the structure factor remains qualitatively unchanged over a large temperature range~\cite{Gardner2001,Castelnovo2019}: not only for $T \gg |\mathcal{J}|$ (where it is controlled by $\langle S_i S_j \rangle \sim -\mathcal{J}_{ij}/T$, namely the leading order term in a $|\mathcal{J}|/T$ expansion) but also down to temperatures well below the Curie-Weiss temperature, which quantifies the strength of magnetic interactions $|\mathcal{J}|$. In this so-called cooperative paramagnetic regime~\cite{Villain1979}, correlations remain largely short-ranged.

In the present paper, we investigate a natural classification of the classical spin states for a single-tetrahedron Hamiltonian's couplings $\mathcal{J}$ regarded as a $12 \times 12$ matrix. By studying the eigenspaces of this matrix, and spin correlators as projectors onto them, we explore the ground states and the correspondence between high-temperature and $T = 0$ single-tetrahedron correlators in detail. This correspondence is shown to imply the intermediate-temperature regime mentioned above. Finally, we discuss how our results are modified in the presence of phase coexistence for parameter values proximal to a phase transition, as is the case for Yb$_2$Ti$_2$O$_7$. 
%
%

\section{Eigenspaces of the coupling matrix}\label{sec:2}

Following the convention in Ref.~\cite{Ross2011} (see also Appendix~\ref{app:notation}), the generic pyrochlore Hamiltonian can be written as 
\begin{equation}
    \begin{split}
        H = \sum_{\langle Ia, Jb\rangle} 
    &\Big[ 
        J_{zz} S_{Ia}^z S_{Jb}^z -J_{\pm} (S_{Ia}^+ S_{Jb}^- + S_{Ia} ^- S_{Jb}^+) 
        \\ 
        &+ J_{\pm \pm} (\gamma_{ab} S_{Ia}^+ S_{Jb}^+ + \gamma_{ab}^* S_{Ia}^- S_{Jb}^-) 
        \\
        &+ J_{z\pm} \big\{ S_{Ia}^z(\zeta_{ab}S_{Jb}^+ + \zeta_{ab}^* S_{Jb}^- ) + Ia\leftrightarrow Jb\big\} \Big]
        \, ,
    \end{split}
\label{eq:genHam}
\end{equation}
where 
\begin{equation} 
    \zeta = 
    \begin{pmatrix}
    0 & -1 & e^{i \pi/3} & e^{-i \pi/3} \\
    -1 & 0 & e^{-i \pi/3} & e^{i \pi/3} \\
    e^{i \pi/3} & e^{-i \pi/3} & 0 & -1 \\
    e^{-i \pi/3} & e^{i \pi/3} & -1 & 0
    \end{pmatrix}
    \, , \quad \gamma = - \zeta^*
    \, .
\nonumber 
\end{equation}

For convenience we use here the customary notation for pyrochlore Hamiltonians; however, $(S^x, S^y, S^z)$ in our study is a real classical vector of fixed length, defining $S^{\pm} = S^x \pm i S^y$. The indices $I, J$ label tetrahedra of the pyrochlore lattice, and $a, b \in \{0,1,2,3\}$ label the four spins within a tetrahedron (see Appendix~\ref{app:notation} for details).

Consider now the same Hamiltonian restricted to a single tetrahedron, coupling every pair of spins therein:
\begin{equation}
    H =  \sum_{ ab, \alpha \beta} \mathcal{J}_{ab}^{\alpha \beta} S_a^{\alpha} S_b^{\beta} \equiv  \sum_{i j }  \mathcal{J}_{ij}S_i S_j = \textbf{S}^T \mathcal{J} \textbf{S} 
    \, .
\end{equation}
We wrote here the four coupling constants of Eq.~\eqref{eq:genHam} as an interaction tensor $\mathcal{J}_{ab}^{\alpha\beta}$. The first equality absorbs the site and spin component indices into a single index ($a, \alpha \equiv i$, etc.), and the second frames the expression in terms of the matrix $\mathcal{J}$ acting on the $12$-component vector $\textbf{S}$ (four sites and $x,y,z$ on each site). The correlator for a given spin configuration, $S_a^{\alpha} S_b^{\beta}$, may be seen as the projector $S_i S_j = (\textbf{S}\textbf{S}^{\intercal})_{ij}$. The $a=b$ cases --- four $3 \times 3$ matrices along the block diagonal of the $\textbf{S}\textbf{S}^{\intercal}$ matrix --- are immaterial for what follows. Many equalities/proportionalities hereafter will be considered as being ``modulo'' (i.e., ignoring) these four blocks, and we will use the symbols $\offeq$ / $\offprop$ respectively for these.

Note that a generic vector cannot be translated into a spin configuration, unless the four spins that comprise it have the same length. We henceforth call this ``strong normalisation'', as opposed to the usual, more permissive (``weak'') normalisation, which only fixes the length of the entire 12-dimensional vector. 

\subsection{Classification of single-tetrahedron phases as eigenspaces}\label{sec:2a}

We then look at the eigensystem of $\mathcal{J}$. Say the distinct eigenvalues are $E_0 < E_1 < E_2 < \cdots$ and the corresponding orthonormal eigenvectors are $\textbf{v}_n^{(m)}$, $\mathcal{J}\textbf{v}_n^{(m)} = E_n \textbf{v}_n^{(m)}$, so that $m$ enumerates the eigenvectors within an eigenspace labelled by $n$. If we write the projector on the $n^{\text{th}}$ eigenspace as $P_n = \sum_m \textbf{v}_n^{(m)} (\textbf{v}_n^{(m)})^{\intercal}$, then $\mathcal{J} = \sum_n E_n P_n$.

Using representation theory~\cite{McClarty2009, Yan2017}, one finds that there are 12 orthogonal strongly normalised eigenvectors, which are basis vectors for at most five eigenspaces, constituting bases for irreducible representations of $T_d$. The heuristic argument for this is the following: given a weakly normalised eigenvector with the length of spin $a$ greater than the length of spin $b$, we must be able to perform a symmetry operation that swaps the two spin lengths. This is an eigenvector with the same eigenvalue, and we can then form a superposition of the two where spins $a$ and $b$ are strongly normalised, etc. The set of strongly normalised vectors thus generated will automatically be a basis for an irreducible representation of $T_d$. 
A corollary of this result is that one is guaranteed to find the single-tetrahedron ground states (which must satisfy the strong normalisation condition) within the space with eigenvalue $E_0$. 

The eigenspaces, and the labels we use for the corresponding bases, are the following~\footnote{More details on the eigensystem may be found in Ref.~\cite{Yan2017}, in particular Sections II and III. The eigenvalues which we call $E$ are called $a$ there, and expectation values of projectors $\textbf{S}^{\intercal} P \textbf{S}$ are squares of order parameters, $\textbf{m}^2$.}:

    (1) $A_2$ (one-dimensional): the 4out (or 4in, since this is related by an overall minus sign) state.
   
    (2) $E$ (two-dimensional): the ground states of the $XY$ ferromagnet. A conventional basis is given by the local $\textbf{x}$ and $\textbf{y}$ vectors at each site, and this trivially generates a U(1) easy-plane degeneracy, $\textbf{S}_a= \textbf{x}_a\cos\phi+ \textbf{y}_a\sin\phi $.
    
    (3) $T_2$ (three-dimensional): the Palmer-Chalker states~\cite{Palmer2000}, which are also in the easy planes. 
    
    (4) $T_{1A'}$ (three-dimensional): the splayed phase, forming one basis for the representation $T_1$. The eigenvectors change continuously with the splay angle $\theta_{T_1}$, which is a function of the parameters $\{J_{zz}, J_{\pm}, J_{\pm\pm}, J_{z\pm}\}$. If $J_{z\pm} = 0$, the splay angle becomes independent of the coupling parameters, and so do the eigenvectors, forming three more easy plane states.
    
    (5) $T_{1B'}$ (three-dimensional): these are again functions of the same single parameter $\theta_{T_1}$ (but the functions are distinct from $T_{1A'}$) constituting another basis for $T_1$. If $J_{z\pm} = 0$, they are in fact the three linearly independent 2in-2out states.

Strongly normalised vectors in the lowest eigenspace are ground states, and when the lowest eigenspace changes, so does the set of ground states. At phase boundaries, two eigenvalues coinciding can result in a degeneracy protected by an enhanced symmetry (unstable to generic perturbations) not present in the constituent eigenspaces themselves. A notable example is the ground-state manifold of the Heisenberg ferromagnet in local spin variables ($J_{\pm} = - J_{zz}/2 > 0$, $J_{\pm\pm}=J_{z\pm}=0$)~\footnote{Not to be confused with the global Heisenberg model, which one obtains for $(J_{zz},J_{\pm},J_{\pm\pm},J_{z\pm}) \propto (-1/3, 1/6, 1/3, \sqrt{2}/3)$.}, where the $XY$ ferromagnet and 4out eigenvalues coincide ($E_0 = E_{E} = E_{A_2}$) giving rise to an O(3) degeneracy. All three of the other eigenspaces are also degenerate for this choice of parameters, and they are a basis for the ground-state manifold of the Heisenberg antiferromagnet ($J_{\pm} = - J_{zz}/2 < 0$, $J_{\pm\pm}=J_{z\pm}=0$).


\begin{figure}[t!]
\includegraphics[width=0.23\textwidth]{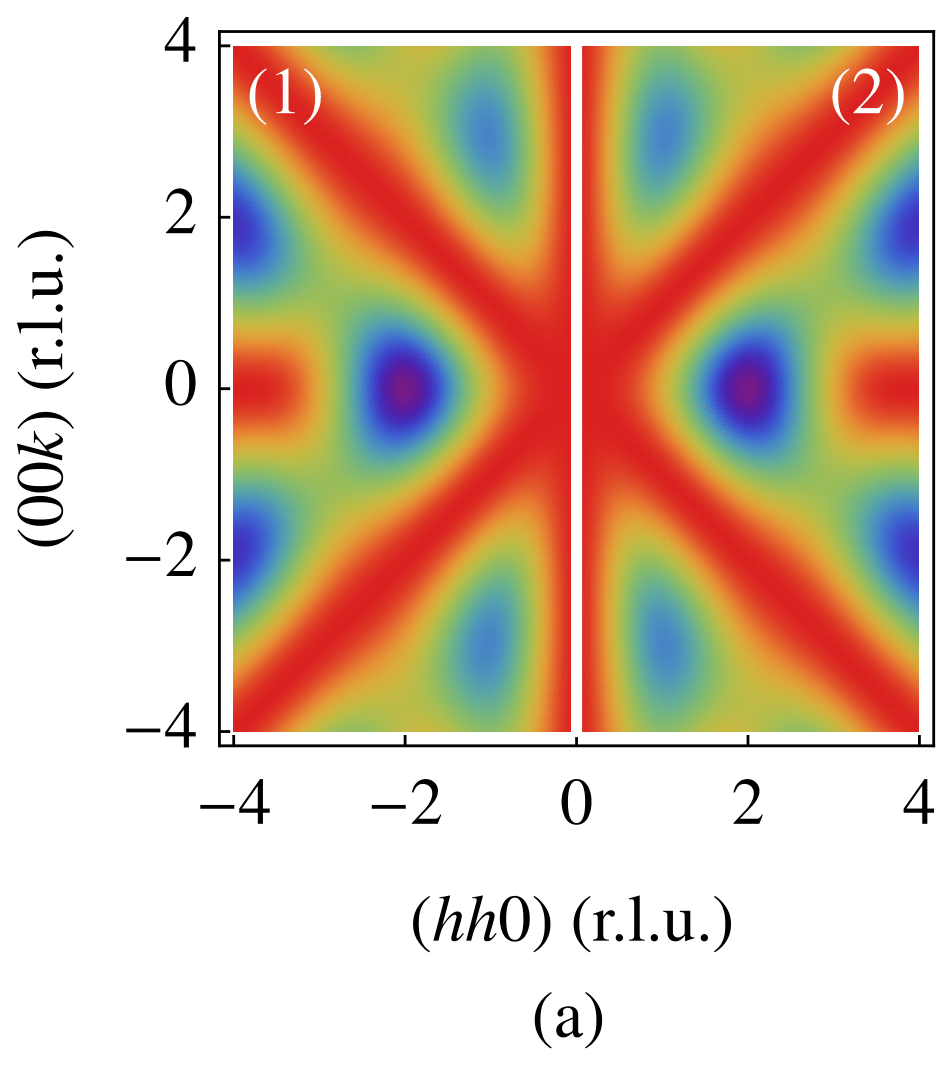}
\includegraphics[width=0.23\textwidth]{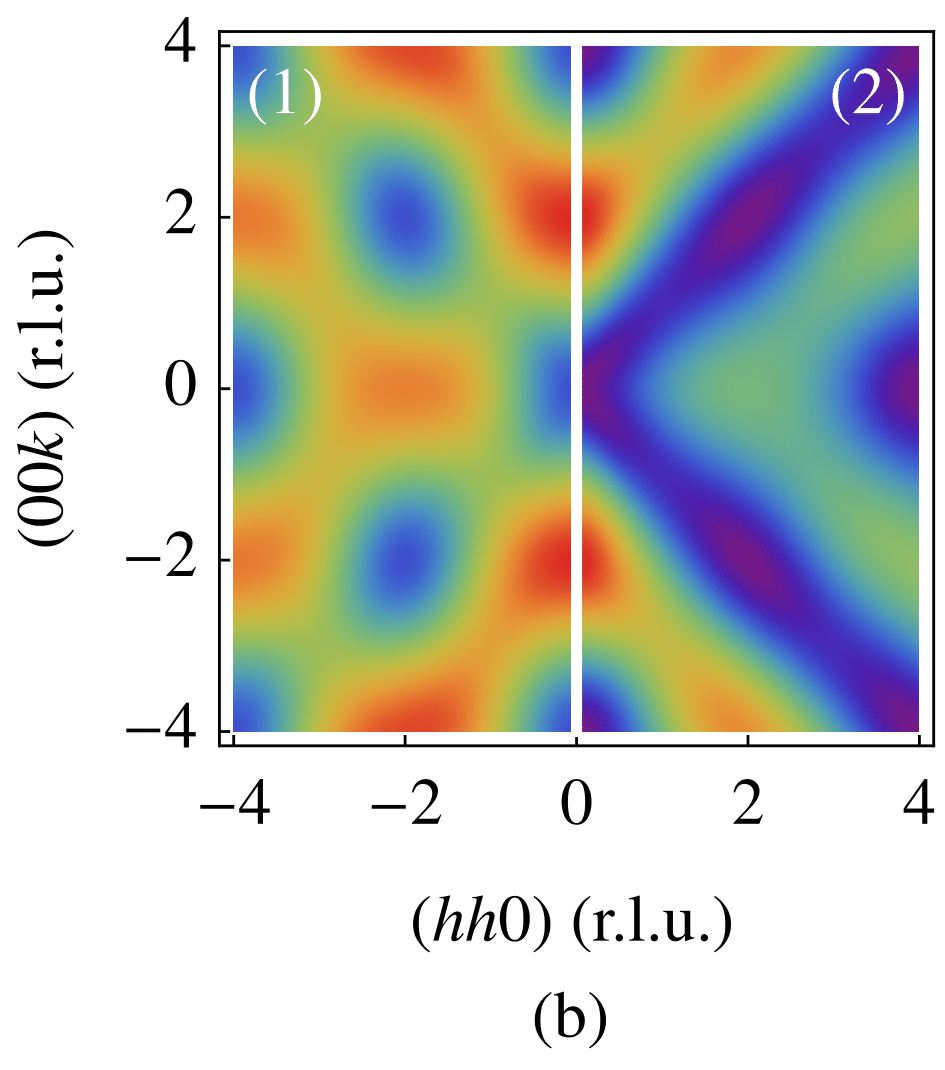}
\caption{The single-tetrahedron structure factor derived from $\langle\textbf{S}\textbf{S}^{\intercal} \rangle \sim -\mathcal{J}$ and $\langle\textbf{S}\textbf{S}^{\intercal} \rangle \sim P_0$ (left and right halves, 1 and 2 respectively, within each panel). 
(a) $(J_{zz}, J_{\pm}, J_{\pm\pm}, J_{z\pm}) = (1,0,0,0)$: the $P_0 \offprop -\mathcal{J}$ correspondence holds, and the two structure factors are exactly proportional to one another. 
(b) $(J_{zz}, J_{\pm}, J_{\pm\pm}, J_{z\pm}) = (0,0,0,1)$: the $P_0 \offprop -\mathcal{J}$ correspondence does not hold, and the two structure factors are obviously different. However, a certain degree of qualitative similarity is still apparent.}
\label{fig:jzz_jzpm} 
\end{figure} 


\subsection{Correlators at high $T$ and $T = 0$}\label{sec:2b}

Consider the spin correlator at temperature $T$: 
\begin{equation}
    \langle S_i S_j \rangle \propto \sum_{\{S\}} S_i S_j \exp\Bigg(-\frac{1}{T}\sum_{kl} \mathcal{J}_{kl} S_k S_l\Bigg)
    \, ,
\end{equation}
where the indices are no longer restricted to a single tetrahedron, and the sum on $\{S\}$ is over all possible spin configurations. Expanding to first order in $1/T$, one obtains:
\begin{equation}\label{eqn:expansion}
    \langle S_i S_j \rangle \sim \sum_{\{S\}} S_i S_j - \frac{1}{T}\sum_{\{S\}}\sum_{kl} \mathcal{J}_{kl} S_i S_j S_k S_l + \cdots 
    \, . 
\end{equation}
The zeroth order vanishes, since $S_i$ and $S_j$ independently average to zero ($i \neq j$). The $1/T$ contribution is proportional to $-\mathcal{J}_{ij}$, as only terms of the form $\mathcal{J}_{ij} S_i^2 S_j^2$ survive. The leading order correlations at high $T$ are therefore $\langle S_i S_j \rangle \sim - \mathcal{J}_{ij}/T$ if $i$ and $j$ are nearest neighbours, and zero otherwise.
The high-temperature single-tetrahedron correlator can then be written as a matrix:
\begin{equation}
    \langle S_i S_j \rangle \sim -\frac{\mathcal{J}_{ij}}{T} = -\frac{1}{T} \sum_n E_n (P_n)_{ij} 
    \, .
\end{equation}

To obtain the ground-state correlator of a single tetrahedron, one needs to take the limit $T \rightarrow 0^+$ with care to account for potential order-by-disorder effects. One case where the answer is clear is when the $P_0 \offprop -\mathcal{J}$ correspondence (discussed in the following sections) holds; as we shall show, here $\langle\textbf{S}\textbf{S}^{\intercal} \rangle \offprop -\mathcal{J} \offprop P_0$ at all $T > 0$, and hence also in the $T \rightarrow 0^+$ limit.
For Hamiltonians that have a finite number of strongly normalised ground states, $P_0$ being the $T=0$ correlator remains a reasonable proposition on the grounds that the eigenstates $\textbf{v}_0^{(m)}$ are related by symmetries of the tetrahedron, and therefore their correlators should be weighted equally, yielding $\langle\textbf{S}\textbf{S}^{\intercal} \rangle \offprop \sum_m \textbf{v}_0^{(m)} (\textbf{v}_0^{(m )})^{\intercal} = P_0$. We verified this via Monte Carlo simulations on several instances of such Hamiltonians. 
The case of continuously many strongly normalised ground states (without $P_0 \offprop \mathcal{J}$) is more difficult, but there is only one possibility where such continuity does not rely on a fine-tuned degeneracy between irreps at eigenvalue $E_0$: the U(1) manifold of the irrep $E$, $\textbf{S}_a= \textbf{x}_a\cos\phi+ \textbf{y}_a\sin\phi $. The correlator here is still proportional to $\textbf{x}\textbf{x}^{\intercal} + \textbf{y}\textbf{y}^{\intercal} = P_E$; this is again confirmed by Monte Carlo simulations 
\footnote{Moreover, we can argue that even if the $T \rightarrow 0^+$ limit entropically prefers some $\phi$ over others, a $\mathbb{Z}_N$ symmetry $\phi \mapsto \phi + 2 \pi /N$ must survive --- see Sec.~\ref{sec:ybto} for an example of $N=6$ --- so that the averaged correlator remains $P_E$.}. 
However, it is not always true that the $T=0$ correlator is $\langle\textbf{S}\textbf{S}^{\intercal} \rangle \offprop P_0$, since order by disorder can be important on a (measure zero) set of fine-tuned points in parameter space where geometrically complex ground state manifolds appear.

\subsection{High $T$ versus $T=0$ correspondence}\label{subsec:corr}

With this language in place, we formulate the central statement of this work: the high-temperature correlator is proportional to the $T = 0$ single-tetrahedron one for coupling matrices $\mathcal{J}$ where $P_0 \offprop -\mathcal{J}$. At a few high symmetry points in parameter space this relation holds exactly, whereas a close resemblance appears to hold quite generally. Figures \ref{fig:jzz_jzpm}(a) and \ref{fig:jzz_jzpm}(b) show examples of parameters for which the correspondence does and does not hold, respectively. Generically, $\mathcal{J}$ can have five distinct eigenvalues. When $P_0 \offprop -\mathcal{J}$ exactly, we observe that there are either two or three distinct eigenvalues, and we comment on the two cases separately.

If there are two eigenvalues, we know that $P_0 + P_1 = I$ by completeness. Given that the four self-correlating ($a=b$) $3 \times 3$ blocks along the diagonal are irrelevant, this equation implies that the ``physical'' entries in the projectors satisfy $P_1 \offeq -P_0$. The decomposition $\mathcal{J} = E_0 P_0 + E_1 P_1$ therefore reduces to $\mathcal{J} \offeq (E_0 - E_1) P_0 $ and thus $P_0 \offprop -\mathcal{J}$ (in our convention $E_1 > E_0$). This only happens for the two Heisenberg Hamiltonians (local/global axes).

In the case of three eigenvalues, if any one projector is a multiple of the other, then the third is automatically also a multiple due to completeness: say if $P_2 \offeq a P_0$ ($a \neq 0$), then $P_0 + P_1 + a P_0 \offeq 0$, and $P_1 \offeq -(1+a)P_0$. Therefore, $\mathcal{J} = E_0 P_0 + E_1 P_1 + E_2 P_2 \offeq [(E_0 - E_1) + a (E_2 - E_1)]P_0$, and $E_0 < E_1 < E_2$ implies that if $a < 0$, then $P_0 \offprop - \mathcal{J}$. Conversely, the only way for $P_0 \offprop - \mathcal{J}$ to hold is if all three projectors are proportional to each other. Furthermore, we can show that $a=-1$, i.e. $P_0 \offeq -P_2$ and $P_1 \offeq 0$. This proof is presented for convenience in Appendix~\ref{app:proofs}.

It is interesting to compare this result to the case of spin ice in Ref.~\cite{Castelnovo2019}. The apparent similarity is deceiving, as we explain for convenience in Appendix~\ref{app:monopoles}: the set of configurations with a vanishing correlator is inherently different in the two cases. 

We now explain why the correlators for $\mathcal{J}$ which obey the correspondence behave as $\langle S_i S_j \rangle \propto -\mathcal{J}_{ij}$ already at intermediate temperatures. Consider a single tetrahedron at arbitrary temperature $T$. The coefficient of $1/T^n$ in Eq.~\eqref{eqn:expansion} contains all possible contractions from $i$ to $j$ connected by $n$ instances of $\mathcal{J}$. Translating back to matrices, the only possible objects that can result are powers $\mathcal{J}^n$; all other possible terms vanish due to either a vanishing self-coupling of the form $\mathcal{J}_{ii}$, or an unpaired $S$ that averages to zero. For instance, the only non-trivial terms at order $1/T^2$ are of the form $\mathcal{J}_{ik}\mathcal{J}_{kj} S_i^2 S_j^2 S_k^2$. If $\mathcal{J}$ is then proportional to $P_0$, we have that $\mathcal{J}^n \offprop \mathcal{J}$. This last relation is proven in Appendix~\ref{app:proofs}.

Therefore, if the correspondence holds, then a single tetrahedron has the same correlator at all $T$, up to a multiplicative factor which is a function of $T$. Finally, at temperatures high enough that the correlations of a thermodynamically large system vanish rapidly beyond nearest-neighbour distances, one can reasonably approximate the structure factor of the system with that of a single tetrahedron at the same temperature, and thence with that of $P_0$. 
%
%
\begin{figure}[b!]
\includegraphics[width=0.4\textwidth]{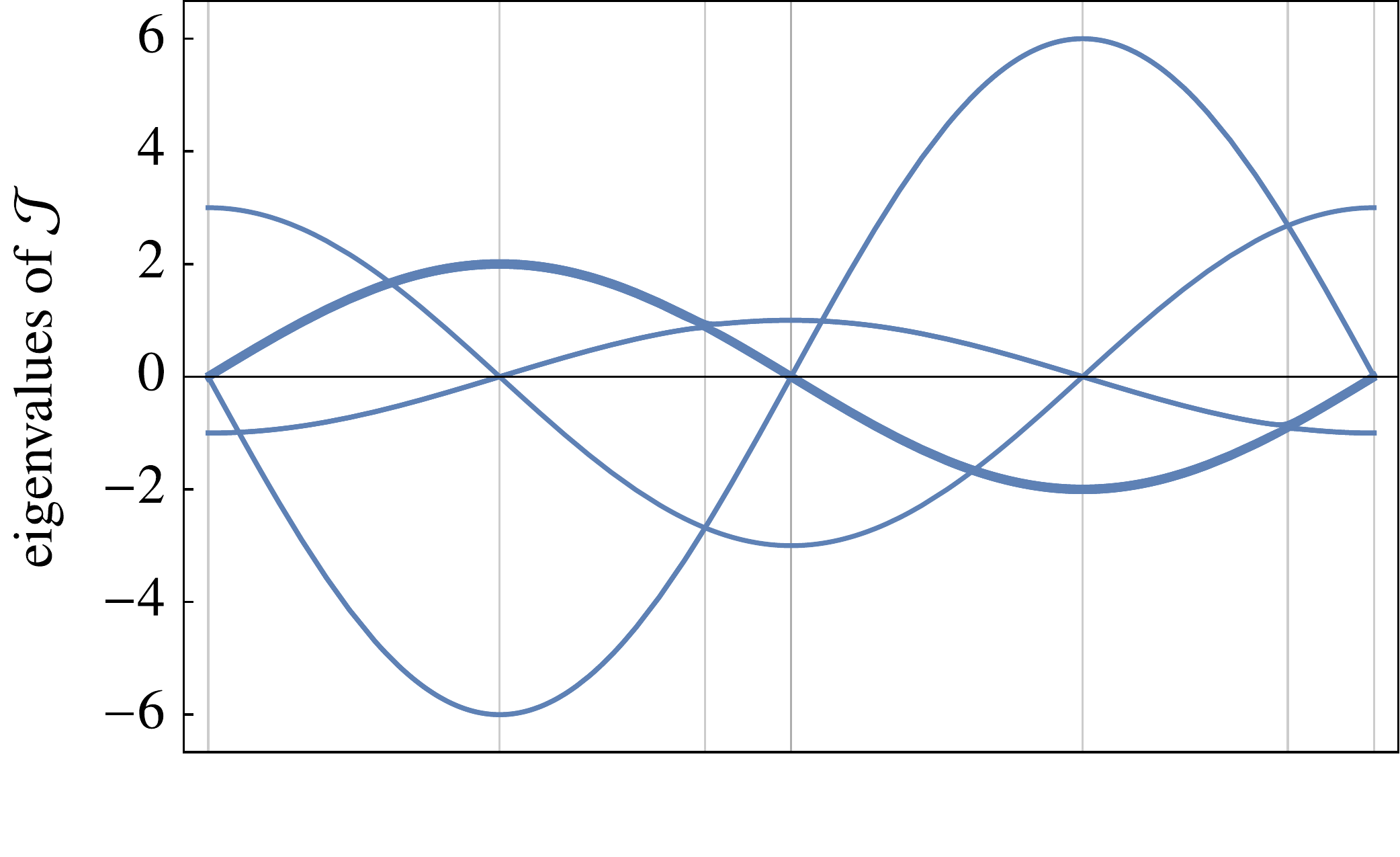}
\includegraphics[width=0.4\textwidth]{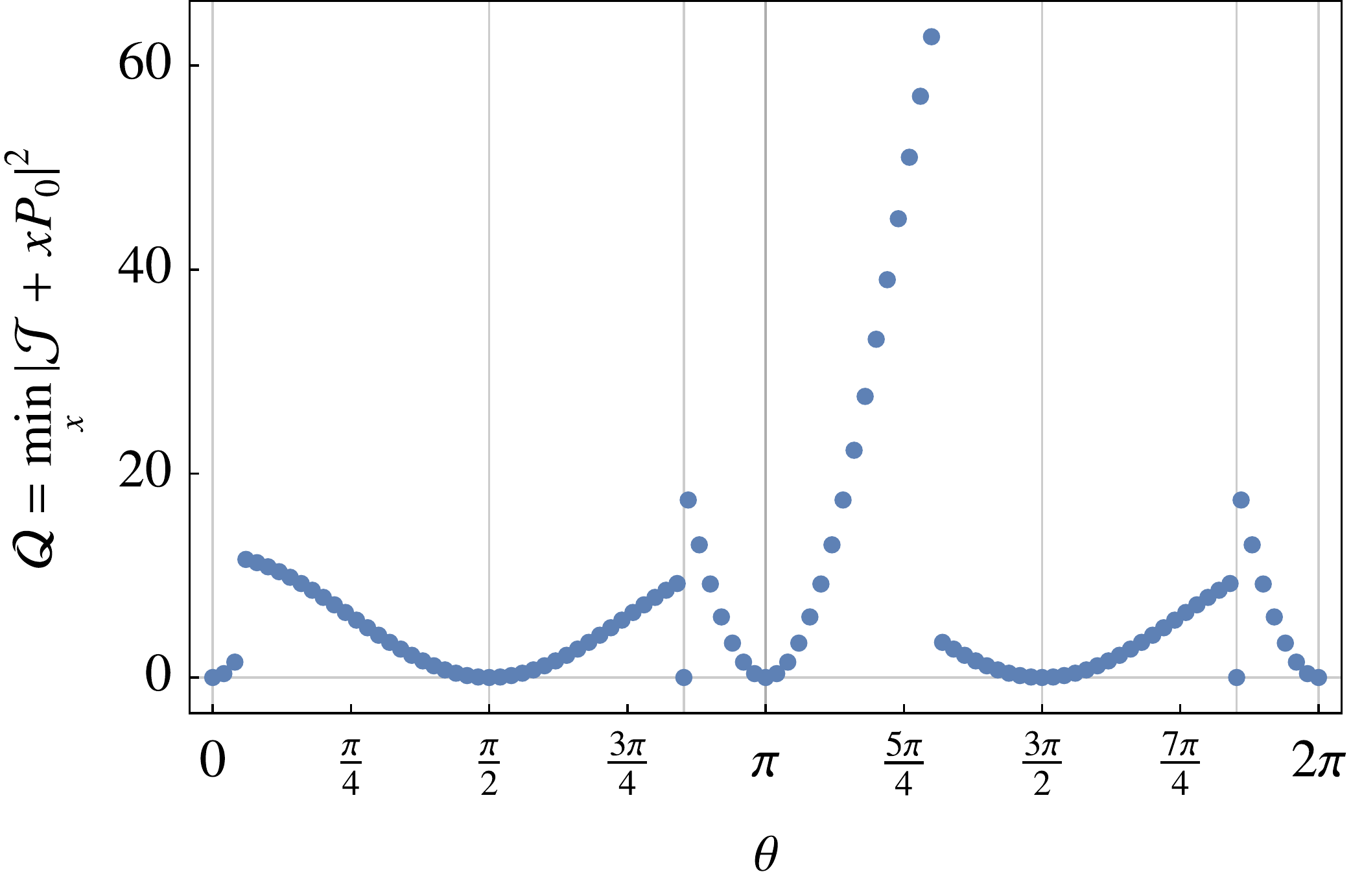}
\caption{Eigenvalues of $\mathcal{J}$ and deviation $\mathcal{Q}$ from $\mathcal{J} \offprop -P_0$ (top and bottom panel, respectively), for $(J_{zz}, J_{\pm}, J_{\pm\pm}, J_{z\pm}) = (\cos \theta, \sin \theta, 0,0)$ against $\theta \in [0, 2 \pi].$ Vertical gridlines are shown at points where $\mathcal{J} \offprop -P_0$ holds exactly, i.e. where $\mathcal{Q}=0$. One of the eigenvalues, shown as a thicker line, contains two eigenspaces, $T_2$ and $T_{1A'}$.}
\label{fig:xxz} 
\end{figure}

\begin{figure*}[ht!]
\includegraphics[width=0.23\textwidth]{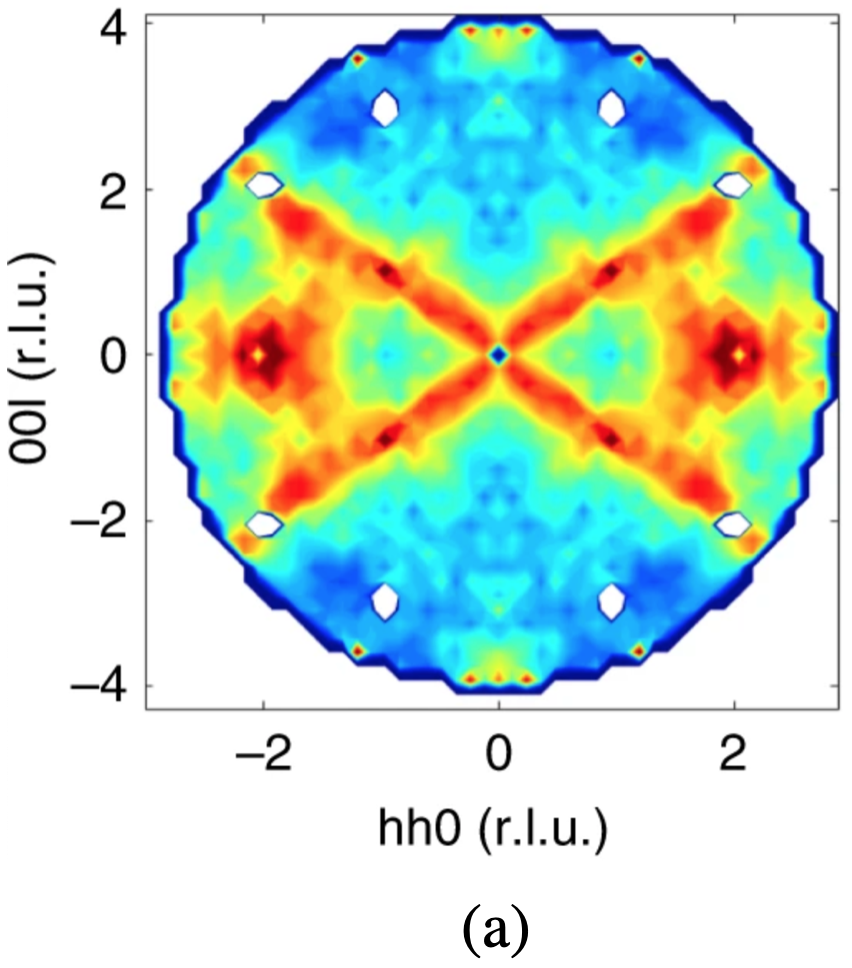}
\includegraphics[width=0.23\textwidth]{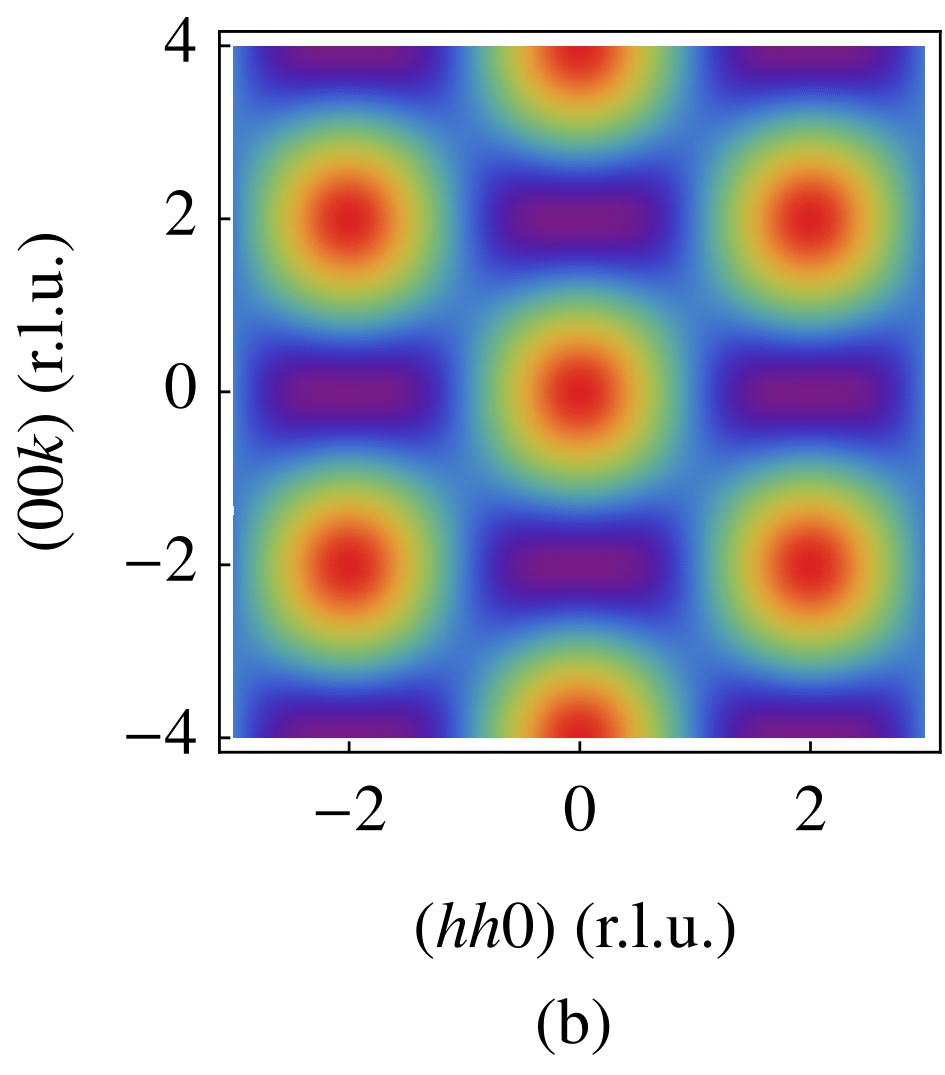}
\includegraphics[width=0.23\textwidth]{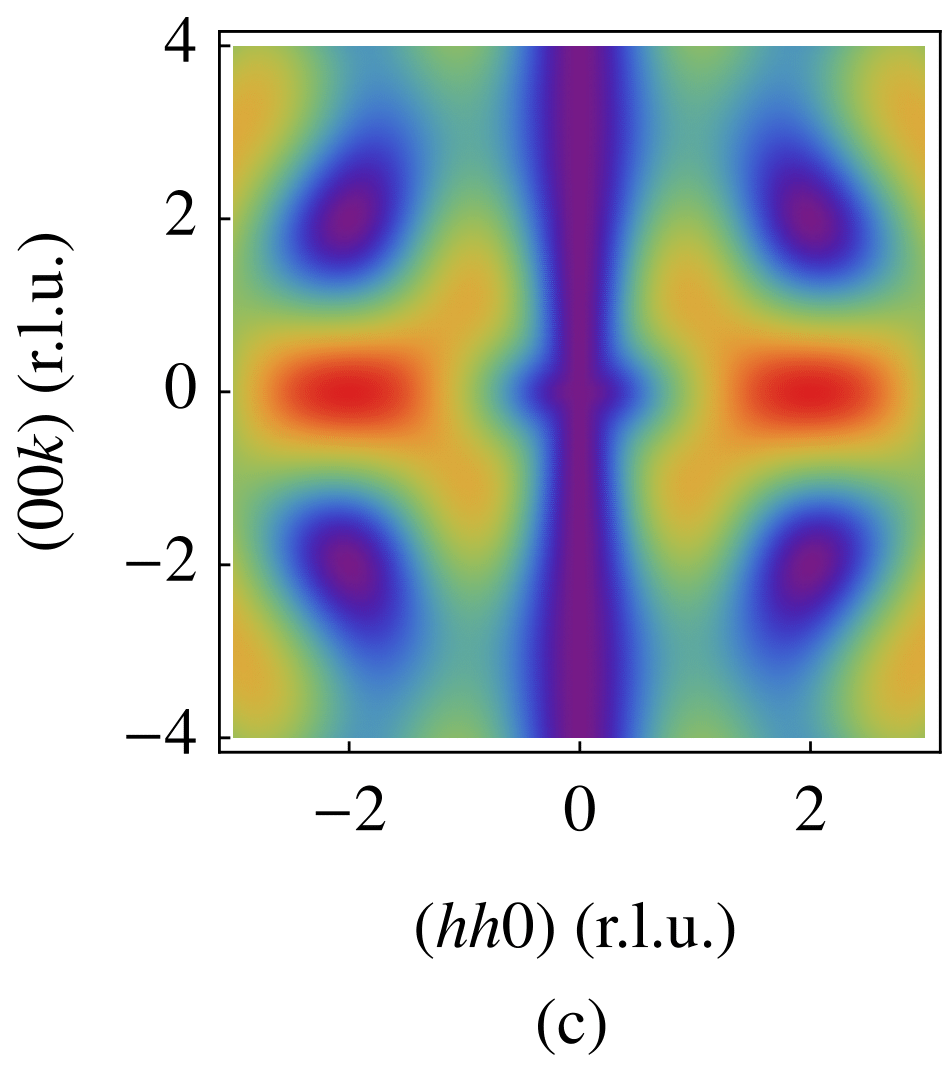}
\includegraphics[width=0.23\textwidth]{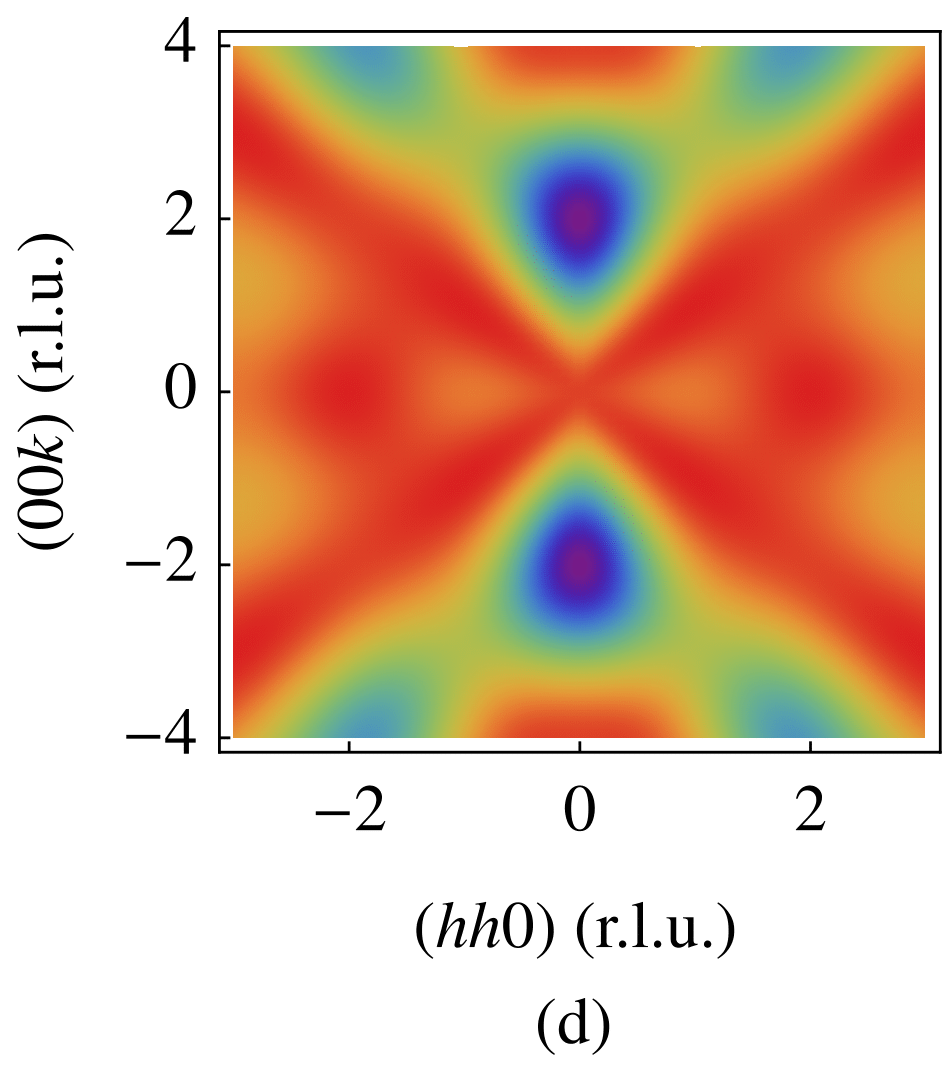}
\caption{(a) Fig. 4a of Ref.~\cite{Bowman2019}, showing the experimental structure factor of Yb$_2$Ti$_2$O$_7$ at $T \approx 50$~mK; single-tetrahedron structure factors from (b) $\langle \textbf{S}\textbf{S}^{\intercal} \rangle \sim  P_{T_{1A'}}$, (c) $\langle \textbf{S}\textbf{S}^{\intercal} \rangle \sim P_{E}  $, and (d) $\langle \textbf{S}\textbf{S}^{\intercal}\rangle \sim P_{T_{1A'}}+(3/2) P_{E} $. All plots use the $g$ factor $g_{xy}/g_z = 2$.}
\label{fig:ybto} 
\end{figure*} 

\subsection{Examples} \label{sec:2d}

Let us introduce a heuristic measure for the deviation from the correspondence $\mathcal{J} \offprop -P_0$, defined as
\begin{equation}
    \mathcal{Q} = \min_{x \in \mathbb{R}} |\mathcal{J} + x P_0|^2
    \, , 
\end{equation}
where we have chosen the matrix norm to be $|M|^2 = \sum_{a \neq b, \alpha \beta} |M_{ab}^{\alpha \beta}|^2$, i.e., the sum of the squares of all elements that are not self-couplings. Notice that, under the trivial rescaling $\mathcal{J} \mapsto \alpha \mathcal{J}$ (with $\alpha > 0$), $\mathcal{Q}$ scales as $\mathcal{Q} \mapsto \alpha^2 \mathcal{Q}$.  As an example, the spectrum of $\mathcal{J}$ and the behaviour of $\mathcal{Q}$ for $(J_{zz}, J_{\pm}, J_{\pm\pm}, J_{z\pm}) = (\cos \theta, \sin \theta, 0,0)$, as a function of $\theta \in [0, 2 \pi]$, are shown in Fig.~\ref{fig:xxz}. Restricting parameters to this circle is sufficient to span the entire space of $XXZ$ Hamiltonians, up to the aforementioned trivial overall scaling. The exact definition of $\mathcal{Q}$ remains somewhat arbitrary, but this is unimportant for the features we discuss. Within this manifold, the four eigenspaces belong to $T_{1B'}, E, T_2+T_{1A'}$ (degenerate), and $A_2$; for small positive $\theta$, these appear in increasing order of eigenvalues. The eigenvectors do not change with $\theta$, as $J_{z\pm} = 0$ is always satisfied. Also note that $P_{T_{1B'}} + P_{A_2} \offeq 0$, since these are the 2in-2out and 4in/4out correlators, respectively. By completeness, also $P_{E} + P_{T_2} + P_{{T_{1A'}}}\offeq 0$. Therefore, the correspondence holds ($\mathcal{Q}=0$) when there are two eigenvalues (``local'' Heisenberg: see in Fig.~\ref{fig:xxz} the points between $3\pi/4 $ and $\pi$, and between $7\pi/4$ and $2\pi$), or three with a crossing of $T_{1B'}$ and $A_2$ (purely $J_{\pm}$: $\theta = \pi/2, 3\pi/2$) or $E$ and $T_2+T_{1A'}$ (purely $J_{zz}$: $\theta = 0, \pi$).

When there is a level crossing, we have exactly one of the following: 

   (1) The deviation $\mathcal{Q}$ jumps discontinuously from one nonzero value to another. This signals that $P_0$ has changed, and it happens when the lowest and second-lowest eigenspaces swap roles due to $E_0$ and $E_1$ crossing. However, the total projector $P_0+P_1$ at the crossing is not proportional to $-\mathcal{J}$, as opposed to item (3) in this list.
   
   (2) $\mathcal{Q}$ goes to zero continuously, along trajectories where $P_0$ remains unchanged, but higher eigenvalues cross so as to result in the three eigenvalue possibility detailed above.
   
   (3) $\mathcal{Q}$ jumps, and at precisely the location of the discontinuity, it goes to zero. Here there is a lowest-level crossing, and $P_0 + P_1 \offprop -\mathcal{J}$ at the crossing.
   
   (4) $\mathcal{Q}$ neither jumps nor vanishes. This is when $P_0$ is not involved in the crossing, and the higher level crossing does not result in $P_0 \offprop -\mathcal{J}$, as opposed to item (2).

Even when the correspondence does not hold exactly but the deviation is small --- this generically happens when we are far from jumps, although $\mathcal{Q}$ may spike to varying magnitudes near different jumps as can be seen in Fig.~\ref{fig:xxz} --- we expect the physical consequences of $P_0 \offprop -\mathcal{J}$ to remain approximately valid (see Fig.~\ref{fig:jzz_jzpm}(b)). Moreover, note that the relation $\mathcal{J}^n \offprop \mathcal{J}$, and hence the robustness of correlators in the cooperative paramagnetic regime, can only be continuously violated near a point in parameter space where $P_0 \offprop -\mathcal{J}$ ($\mathcal{Q} =0$) even if $\mathcal{Q}$ jumps from one nonzero value to another in its neighbourhood.

%
%

\section{\boldmath Single-tetrahedron phase competition in Y\lowercase{b}$_2$T\lowercase{i}$_2$O$_7$}\label{sec:ybto}

It is well known that the coupling parameters of Yb$_2$Ti$_2$O$_7$ lies close to the boundary between two classical phases, the splayed ferromagnet, FM ($T_{1A'}$), and the $XY$ ferromagnet ($E$), which is antiferromagnetic in the global axes~\cite{Yan2017, Ross2011, Robert2015, Jaubert2015, Scheie2017, Thompson2017}. Fluctuations select out of the $E$ manifold ($\textbf{S}_a= \textbf{x}_a\cos\phi+ \textbf{y}_a\sin\phi $) a discrete set of ground states dubbed $\Psi_2$ and $\Psi_3$, each comprising six states with $\Psi_2$ at $\phi = n\pi/3$ and $\Psi_3$ at $\phi = (n+1/2)\pi/3$. The FM has six ground states as well, namely plus/minus the three irrep basis vectors. The dominant competition over a large range of temperatures is between FM and either $\Psi_2$ or $\Psi_3$~\cite{Canals2008}.

In Fig.~\ref{fig:ybto}(a), the experimentally determined structure factor at low temperature, plotted in the $(hhk)$ plane, shows two interesting features: rods in the $(111)$ direction~\cite{Thompson2011}, and a maximum at $(220)$ signalling antiferromagnetic correlations in the global frame~\cite{Bowman2019}.  The $T_{1A'}$/$E$ correlators evaluated separately do not yield a satisfactory comparison to the features mentioned above 
(see panels (b) and (c) of Fig.~\ref{fig:ybto}). 

Remarkably, an effective single-tetrahedron model of phase mixing which averages over the $6+6$ ground-state correlators is in fact able to reproduce these characteristic features, as shown in Fig.~\ref{fig:ybto}(d). 
A microscopic justification for this could be that temperatures are high enough for the FM and $\Psi_2/\Psi_3$ phases to be effectively degenerate and correlations to be predominantly nearest-neighbour, while being low enough that the thermal occupation of higher eigenspaces is sufficiently suppressed. In this regime one could argue that fluctuations populate equally the ground states in the two lowest, nearly degenerate eigenspaces, yielding an effective microcanonical picture: the probability of being in a phase is proportional to the number of available states. Thus, if FM and $\Psi_2$ (or $\Psi_3$) mix, the total correlator (taking the numerical factors between projectors and correlators into account) is proportional to $P_{T_{1A'}}+(3/2) P_{E} $. This produces the $(111)$ rods and $(220)$ maximum in the structure factor, as shown in Fig.~\ref{fig:ybto}(d).

However, this is not yet the full story. Performing Monte Carlo simulations in the relevant temperature regime does reproduce the experimental structure factor reasonably well (see Appendix~\ref{app:rods} and also Fig.~4 in Ref.~\cite{Bowman2019}). But if we truncate the spin correlators at nearest-neighbour distance before we compute the structure factor, only the $(111)$ rods appear (see Appendix~\ref{app:rods}). The $(220)$ maximum must therefore come from correlations beyond a single tetrahedron that are only accidentally reproduced by the phase-mixing modelling above. 

%
%

\section{Conclusions and outlook} 

We have shown that intermediate-temperature correlations in pyrochlore magnets can be understood remarkably well via single-tetrahedron ground states. Cases where the correspondence is most precise were discussed, and we further illustrated using parameters appropriate for Yb$_2$Ti$_2$O$_7$ that even when a system lives at the edge between two phases and is not characterised accurately by either at finite temperature, an effective model of the competition can recover important qualitative results. 

Further work in this direction can focus on several aspects:
How does one repeat our prescription for other lattices? In particular, what conditions do we need to find strongly normalised vectors in the lowest eigenspace? This is closely related to the Luttinger-Tisza problem~\cite{Luttinger1946, Litvin1974}. We found parameters for which the correspondence holds by brute force computation, but perhaps there is a more symmetry-motivated systematic approach.

All our results are for vector spins that are free to rotate in three dimensions. One could ask if a similar eigenspace picture can be developed for constrained models, say where the spins are either easy plane or easy axis. This is indeed relevant for pyrochlore materials, where large single-ion crystal-field terms often lead to an anisotropic behaviour of the spins. A natural intermediate object of study could be a model where we only allow strongly normalised eigenvectors as configurations, with energies defined by the same $\mathcal{J}$. For instance, the allowed configurations for a $J_{zz}$ Hamiltonian (whose relation to spin ice we comment on in Appendix~\ref{app:monopoles}) would be ice rule vacua, double monopoles, and easy axis zero modes. Linear combinations of these configurations span all possible states, which makes the phase space at $T>0$ broaden continuously. A discrete constraint avoids this complication, and hence permits a closer analogy to arguments about the temperature dependence of correlators presented in Ref.~\cite{Castelnovo2019}.

Is the kind of naive configuration mixing outlined for Yb$_2$Ti$_2$O$_7$, which has an accidental near-degeneracy in the lowest eigenspaces, perhaps more widely useful as a concept for magnetic materials with strongly frustrated couplings? The observation that long-ranged correlations yield a structure factor that is similar to a short-ranged model is obviously not a precise one in our formulation; can more be said about the microscopics leading to this, or is it truly coincidental? It would also be interesting to see in more detail what other observable consequences a strong phase competition engenders. At any rate, the usefulness of such a simple scheme for relatively complex magnets is remarkable, and it may be worth investigating to what extent one can use this as a starting point for more systematic approximations. 
%
%

\begin{acknowledgements} 

We would like to thank John Chalker, Hitesh Changlani, Michel Gingras, Paul McClarty, Jeff Rau, and Attila Szab\'{o} for useful discussions. We are particularly grateful to one of the Referees for pointing out to us the idea underlying the proof of $a=-1$ in Appendix~\ref{app:proofs}. This work was supported in part by Corpus Christi College, Cambridge and the Leibniz Programme of the DFG (AP); the DFG through SFB 1143 (project id 247310070) and ct.qmat (EXC 2147, project id 39085490), and the National Science Foundation under Grant No. NSF PHY-1748958 (RM); and the Engineering and Physical Sciences Research Council (EPSRC) Grants No.~EP/K028960/1 and  No.~EP/P034616/1 (CC). 

\end{acknowledgements}

\appendix

\section{Notation and parametrisation}\label{app:notation}

The pyrochlore lattice is defined starting with an fcc lattice with a cubic unit cell of side $l$, indexed by $\textbf{R}_I$, and placing a lattice point at each $\textbf{R}_I + \Delta_a$  for all $I$, and $a \in \{0,1,2,3\}$. Here 
\begin{equation}
\begin{split}
    \Delta_0 = \frac{l}{8} (1,1,1) 
    \qquad	
    &\Delta_1 = \frac{l}{8} (1,-1,-1) \\
    \Delta_2 = \frac{l}{8} (-1,1,-1) 
    \qquad
    &\Delta_3 = \frac{l}{8} (-1,-1,1)
\end{split}
\end{equation} 
denote the vertices of a tetrahedron with respect to its centre. The outward pointing unit vectors $\Delta_a/|\Delta_a|$ at each sublattice site define the local $z$ axes. We also define local $x$ and $y$ axes:
\begin{equation}
    \begin{split}
    \textbf{x}_0 &= \frac{1}{\sqrt6}(-2,1,1), \quad 
    \textbf{y}_0 = \frac{1}{\sqrt2}(0,-1,1), 
    \\
    \textbf{x}_1 &= \frac{1}{\sqrt6}(-2,-1,-1), \quad
    \textbf{y}_1 = \frac{1}{\sqrt2}(0,1,-1), 
    \\
    \textbf{x}_2 &= \frac{1}{\sqrt6}(2,1,-1), \quad 
    \textbf{y}_2 = \frac{1}{\sqrt2}(0,-1,-1), 
    \\
    \textbf{x}_3 &= \frac{1}{\sqrt6}(2,-1,1), \quad 
    \textbf{y}_3 = \frac{1}{\sqrt2}(0,1,1)
    \, . 
\end{split}
\end{equation}
The Hamiltonian is defined on effective spin-1/2 variables $\textbf{S}_{Ia}$ 
at each lattice site. These are the lowest energy doublets in the crystal field splitting of the angular momentum multiplet associated with the magnetic ions. In this work, all interactions are assumed to be at the nearest-neighbour level, and the generic Hamiltonian takes the form of Eq.~\eqref{eq:genHam}. 

The central objects of study are the spin correlator $\langle S_{Ia}^{\alpha} S_{Jb}^{\beta}\rangle$ and the elastic neutron scattering structure factor~\cite{jensen} (up to a multiplicative constant, and ignoring the $q$ dependence of the magnetic form factor):
\begin{equation} \label{ns}
        F(\textbf{q}) = \sum_{IJab \alpha \beta} \Bigg[\delta_{\alpha \beta} - \frac{q_{\alpha}q_{\beta}}{\textbf{q}^2}\Bigg] \langle S_{Ia} ^{\alpha} S_{Jb}^{\beta}\rangle e^{i \textbf{q}\cdot (\textbf{r}_{Ia} - \textbf{r}_{Jb})}
        \, . 
\end{equation}

%
%

\begin{figure*}[t!]
\includegraphics[width=0.23\textwidth]{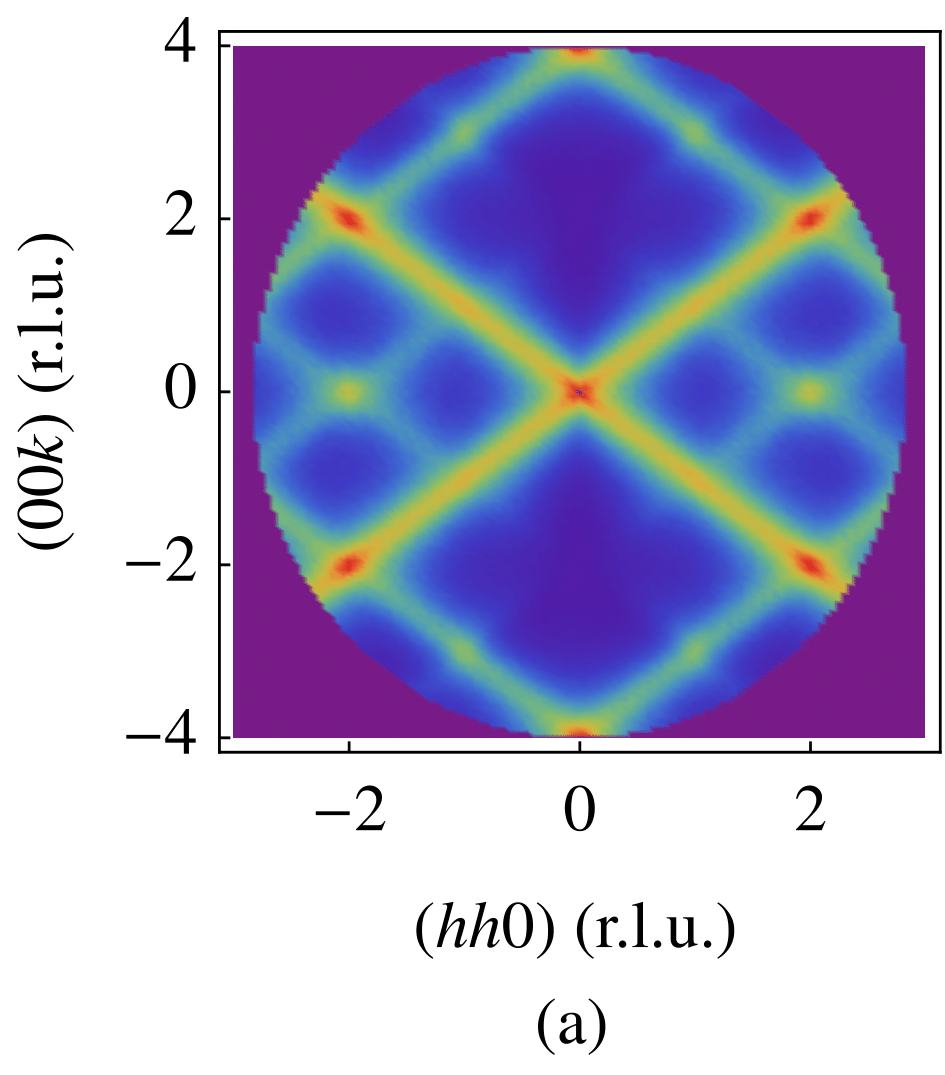}
\includegraphics[width=0.23\textwidth]{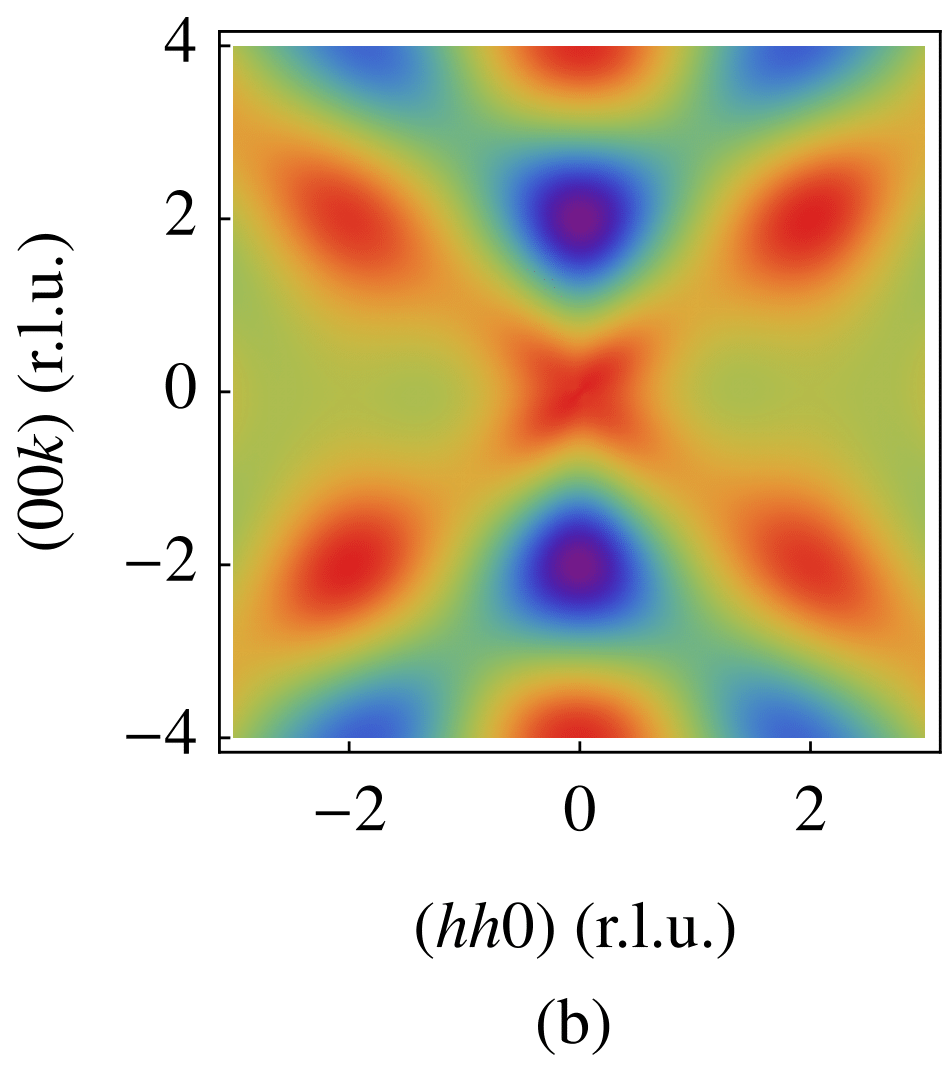}
\includegraphics[width=0.23\textwidth]{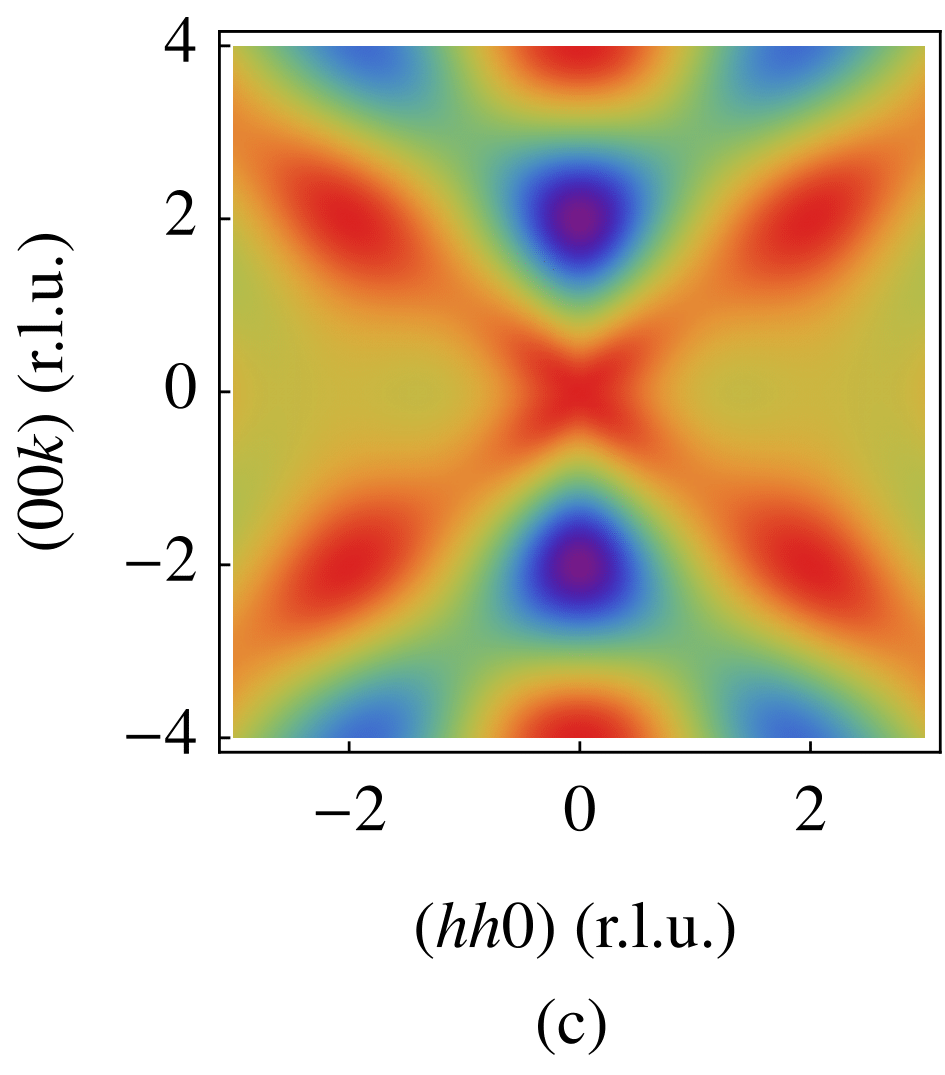}
\caption{Structure factor from Monte Carlo simulations using the Yb$_2$Ti$_2$O$_7$ parameters of Ref.~\cite{Robert2015} and temperature $T = 450$~mK: (a) on a system of $20 \times 20 \times 20$ fcc $16$-spin cubic unit cells and periodic boundary conditions; (b) on the same system, but truncating the correlators to nearest-neighbour distance before computing the structure factor (equivalently, calculating single-tetrahedron structure factors, averaged over all tetrahedra); and (c) on a single isolated tetrahedron.}
\label{fig:ybto_mc} 
\end{figure*}

\section{Proofs of two results in Sec.~\ref{subsec:corr}}\label{app:proofs}

We provide here proofs of two results that require us to retain block-diagonal ($a = b$) terms in the projectors, and therefore we need to abandon ``off'' equalities in favour of conventional ones.

The first is that if we have three eigenvalues and $P_2 \offeq a P_0$ ($a\neq 0$), then $a =-1$. To show this, we write $P_2 - a P_0  = D$, where $D$ is a matrix that only has block-diagonal elements, to wit $D \offeq 0$. Squaring both sides and using the orthogonality of eigenvectors yields $P_2 + a^2 P_0 = D^2$. The difference of these two equations is $(a^2 + a)P_0 = D^2 - D$. If $a^2 + a \neq 0$, $P_0$ is itself block-diagonal. Then $P_2 = a P_0 + D$ is block-diagonal, and $P_1 = I - P_0 - P_2$ is block-diagonal. But $\mathcal{J} = E_0 P_0 + E_1 P_1 + E_2 P_2$ must have off-block-diagonal elements, so this is a contradiction, and $a^2 + a = 0$. Having assumed $a\neq 0$, we have $a = -1$, and $D^2 = D$. These are consistent, since $D = P_2 - aP_0 = P_2 + P_0 = I - P_1$, and indeed $(I - P_1)^2 = I-P_1$. 

The second is that $\mathcal{J} \offprop - P_0$ implies $\mathcal{J}^2 \offprop \mathcal{J}$. The subtlety here is that $\mathcal{J}$ is not exactly a multiple of the projector $P_0$, but also has a block-diagonal part $D'$ which may appear in the physical off-block-diagonal elements of $\mathcal{J}^2$ due to the terms $P_0 D' + D'P_0$. However, we now show that $D'$ is a linear combination of the identity matrix and a different (orthogonal) projector, so that $P_0 D' + D' P_0 \propto P_0$, and thus $\mathcal{J}^2 \offprop \mathcal{J}$ still holds. With two eigenvalues, $\mathcal{J} = E_0 P_0 + E_1 P_1 = E_0 P_0 + E_1 (I - P_0) = (E_0-E_1)P_0+E_1 I$, i.e., $D' = E_1 I$. With three, the arguments of the previous paragraph show that the middle projector $P_1 = I - D$ is block-diagonal. Now, $\mathcal{J} = E_0 P_0 +E_1P_1+E_2P_2 = E_0P_0+E_1P_1+E_2(I-P_0-P_1) = (E_0-E_2)P_0 + E_2 I +(E_1 - E_2)P_1$, and therefore $D' =  E_2 I +(E_1 - E_2)P_1$.

%
%
\section{Comparison to spin ice}\label{app:monopoles}

In classical spin ice, the spins have an Ising character and point either directly into or out of each tetrahedron. The resulting discrete set of tetrahedral configurations can thus be labelled according to the number of incoming vs outgoing spins: 2in-2out; 3in-1out and 3out-1in (also called single monopole configurations); 4in and 4out (double monopoles). 
The nearest-neighbour Hamiltonian reduces to a chemical potential for the monopoles and all the configurations within each label above are isoenergetic. 
Peculiarly, the single monopole configurations yield a vanishing correlator, and therefore the vacuum and double monopole configurations have opposite correlators~\cite{Castelnovo2019}. The analogy with the results presented in the present paper is tempting but deceiving: the set of configurations with a vanishing correlator is in fact different. In our language, the vanishing projector for a $J_{zz}$ Hamiltonian comes from easy plane ``zero modes''. Single monopole configurations are not eigenvectors --- they are superpositions of the vacuum and double monopole eigenvectors such that the average correlator vanishes. 
As we comment in the Conclusions and Outlook, understanding how our results may extend to constrained spin models is an interesting open question. 
%
%

\section{\boldmath Further considerations on Y\lowercase{b}$_2$T\lowercase{i}$_2$O$_7$}\label{app:rods}

We first observe that the experimental structure factor shown in Fig.~\ref{fig:ybto}(a) is nominally at $T \approx 50$~mK, well below the expected ordering temperature of Yb$_2$Ti$_2$O$_7$, $T_c \approx 200$~mK. If the system ordered, we would expect a sharpened version of one of Figs.~\ref{fig:ybto}(b) and \ref{fig:ybto}(c), i.e., either $T_{1A'}$-like or $E$-like correlators. The observed correlators are clearly different, exhibiting characteristic $(111)$ rods and $(220)$ peaks. 

For this reason, we ran Monte Carlo simulations using the parameters of Ref.~\cite{Robert2015} at $T = 450$~mK; a description of the method can be found in Ref.~\cite{Bowman2019}. The structure factor evaluated from $20 \times 20 \times 20$ fcc $16$-spin cubic unit cells and periodic boundary conditions, with $(111)$ rods and $(220)$ maxima, is shown in Fig.~\ref{fig:ybto_mc}(a). We compare this result to the effect of truncating the very same correlators at single-tetrahedron level (i.e., to nearest-neighbour distance) before computing the structure factor (Fig.~\ref{fig:ybto_mc}(b)); and we further compare it to Monte Carlo simulations on a single tetrahedron~\footnote{Of course, single-tetrahedron simulations cannot select the $\Psi_2$/$\Psi_3$ states since they are favoured by an order-by-disorder mechanism that is active only in thermodynamically large systems.} (Fig.~\ref{fig:ybto_mc}(c)). Both approximations only exhibit the $(111)$ rods in the structure factor and not the $(220)$ maxima. 

At closer inspection, both the truncated and single-tetrahedron correlators in our Monte Carlo simulations are in fact very nearly proportional to $P_{T_{1A'}}+P_{E}$ (not shown). This suggests that the single tetrahedron spin states sample uniformly the two lowest eigenspaces~\footnote{Strictly speaking this would include unphysical states that are not strongly normalised, but they can be brought into strongly normalised form by a small admixing of higher excited states.}, which contrasts with sampling only the ground states as in the phase-mixing model in the main text. Moreover, we learn from these results that individual tetrahedra in simulations of thermodynamically large systems have roughly the same correlations as in simulations of a single isolated tetrahedron. 

%
%
\bibliography{b}

\end{document}